\documentclass[conference]{IEEEtran}
\IEEEoverridecommandlockouts
\usepackage{cite}
\usepackage{amsmath,amssymb,amsfonts}
\usepackage{algorithmic}
\usepackage{graphicx}
\usepackage{textcomp}
\usepackage{xcolor}
\def\BibTeX{{\rm B\kern-.05em{\sc i\kern-.025em b}\kern-.08em
    T\kern-.1667em\lower.7ex\hbox{E}\kern-.125emX}}

\usepackage[hidelinks]{hyperref}
\usepackage{stfloats}

\begin{document}

\title{\LARGE \bf Stability Enhancement of Centralized UPS Data Center Systems Under Weak-Grid Conditions \\
\vspace{-0.3cm}


\thanks{\emph{Preprint submitted to IEEE NAPS 2026.} This work has been submitted to the IEEE for possible publication. Copyright may be transferred without notice, after which this version may no longer be accessible.}

\thanks{This manuscript has been authored in part by UT-Battelle, LLC, under contract DE-AC05-00OR22725 with the US Department of Energy (DOE). The US government retains and the publisher, by accepting the work for publication, acknowledges that the US government retains a non-exclusive, paid-up, irrevocable, worldwide license to publish or reproduce the submitted manuscript version of this work, or allow others to do so, for US government purposes. DOE will provide public access to these results of federally sponsored research in accordance with the DOE Public Access Plan.}
}

\newcommand{\orcid}[1]{\href{https://orcid.org/#1}{\textcolor[HTML]{A6CE39}{\aiOrcid}}}


\author{\IEEEauthorblockA{
Jesus D. Vasquez-Plaza$^{\dagger}$\href{https://orcid.org/0000-0001-9514-3124}{\includegraphics[scale=0.12]{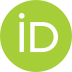}},~\IEEEmembership{(Member,~IEEE)}, Yonghao Gui\href{https://orcid.org/0000-0002-5043-5534}{\includegraphics[scale=0.12]{Figures/orcid.png}},~\IEEEmembership{(Senior Member,~IEEE)},\\ Jin Dong \href{https://orcid.org/0000-0002-5753-1588}{\includegraphics[scale=0.12]{Figures/orcid.png}},~\IEEEmembership{(Senior Member,~IEEE)}, and Jamie Lian\href{https://orcid.org/0000-0003-1270-5350}{\includegraphics[scale=0.12]{Figures/orcid.png}},~\IEEEmembership{(Senior Member,~IEEE)}}
\IEEEauthorblockA{\textit{Oak Ridge National Laboratory}, Oak Ridge, TN 37830, USA}
\IEEEauthorblockA{Email: $^\dagger$vasquezplazj@ornl.gov}
}

\maketitle

\begin{abstract}
Data center power systems are increasingly exposed to weak-grid conditions due to the evolution of modern power systems and the integration of large and dynamic loads. In centralized uninterruptible power supply (UPS) architectures, the front-end rectifier plays a critical role in maintaining stable operation and ensuring reliable power delivery to information technology (IT) equipment. However, conventional phase-locked loop (PLL)-based proportional-integral (PI) control strategies may exhibit degraded performance or instability under low short-circuit ratio (SCR) conditions.
This paper investigates the behavior of centralized UPS systems under weak-grid conditions and demonstrates, through electromagnetic transient simulations, that PI-controlled rectifiers can become unstable at SCR $\approx 2$. To address this issue, a power-based control approach is applied to the three-phase rectifier, enabling direct regulation of active and reactive power without relying on PLL synchronization. 
Simulation results show that the proposed control strategy improves system damping and restores stable operation under weak-grid conditions. The findings highlight the importance of control design for maintaining reliable operation of data center power systems in emerging low-strength grid environments.

\end{abstract}

\begin{IEEEkeywords}
Data centers, AI workloads, phase-locked loop, weak-grids, voltage-modulated direct power control (VM-DPC).
\end{IEEEkeywords}

\section{Introduction}
Data centers have become critical infrastructures in modern power systems due to the rapid growth of cloud computing, artificial intelligence, and large-scale digital services \cite{DataCenter2026}. Their electrical demand is not only high in magnitude, but also increasingly dynamic, as the power consumption of information technology (IT) equipment varies significantly with computational workload and power management strategies \cite{ahmed2021review,chen2509electricity}. Unlike conventional industrial loads, AI-oriented data centers can exhibit fast and synchronized power variations that interact with converter control dynamics and upstream grid conditions.

Centralized uninterruptible power supply (UPS) architectures remain widely used in large-scale data centers \cite{sun2022dynamic,shamseldein2025liability,PNNL2025}. In this topology, the front-end rectifier regulates the dc-link voltage and continuously processes the incoming grid power, supplying a downstream grid-forming inverter (GFM) that feeds the IT load. Therefore, the rectifier is a critical interface between the data center and the grid, and any degradation in its performance can propagate through the dc-link and affect the stability of the overall data center power system~\cite{di2023transforming}. 

Modern power systems are increasingly evolving toward converter-dominated grids with reduced system strength, typically characterized by low short-circuit ratio (SCR). Under such conditions, the interaction between grid impedance and converter control dynamics becomes significant. Conventional PLL-based synchronous reference frame PI control is known to exhibit stability limitations in weak-grids due to impedance interactions, reduced damping, and PLL dynamics \cite{sun2022dynamic,shamseldein2025liability,sun2011impedance,Wang2018,Dong2015,Davari2017,golestan2016three,li2024stability,jeong2022stability}. While these issues have been widely studied for grid-connected converters, they are particularly important for centralized UPS data center systems because the rectifier continuously processes the power required by dynamic IT loads.

To address this challenge, this paper investigates the application of voltage-modulated direct power control (VM-DPC) to the front-end rectifier of a centralized UPS data center system. VM-DPC has been previously proposed for three-phase pulse width modulation (PWM)  rectifiers \cite{Gui2018Rectifier,Gui2021Improved}, HVDC \cite{gil2019direct}, energy storage system \cite{Gong2025Fast}, and weak-grid-connected converters \cite{Gui2019VMDPC}. In this work, the method is applied to a data center UPS architecture subjected to weak-grid conditions and highly dynamic IT load variations. The analysis is carried out using an EMT simulation model that captures the key interactions among the UPS system, the dynamic IT load, and the upstream grid, consistent with recent efforts emphasizing representative data center models for grid-level studies \cite{PNNL2025}.

The main contribution of this paper is to demonstrate that VM-DPC with reactive power support can stabilize a centralized UPS data center system under weak-grid operation, where conventional PI control becomes unstable. The results further show that the stability of the VM-DPC-controlled system depends on the injected reactive power level, highlighting the importance of grid-supportive UPS control strategies for future data center applications.

 The remainder of this paper is organized as follows. Section II describes the centralized UPS data center model and weak-grid representation. Section III presents the rectifier control strategies. Section IV discusses the simulation results, and Section V concludes the paper.

\begin{figure*}
    \centering
    \includegraphics[trim={0.5cm 0.8cm 0.5cm 0.5cm},clip,width=0.85\textwidth]{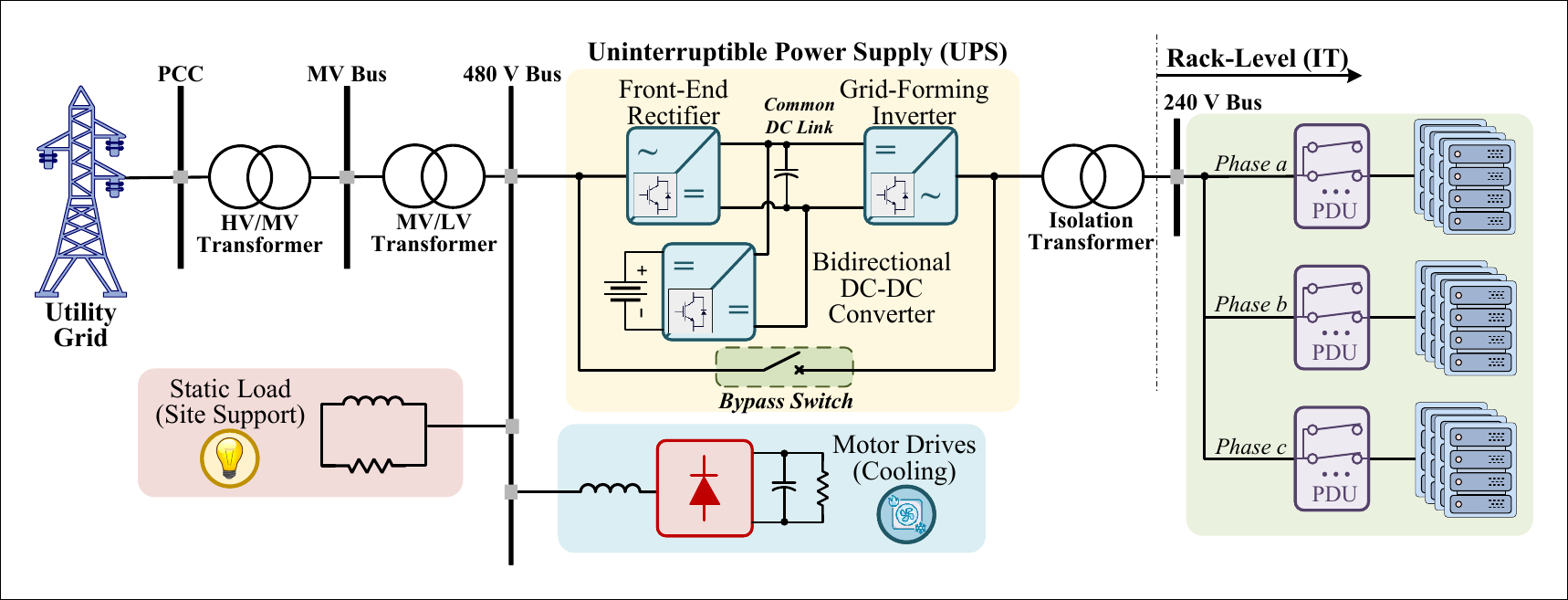}
    \vspace{-1em}
    \caption{Simplified centralized UPS data center architecture.}
    \label{fig:architecture}
    \vspace{-0.3cm}
\end{figure*}

\section{Data Center UPS System Description}

The system considered in this work represents a centralized UPS architecture supplying a large-scale data center.

\subsection{System Architecture}
 As shown in Fig.~\ref{fig:architecture}, the data center is interfaced with the grid through a front-end rectifier, which feeds a dc-link that supplies a GFM responsible for delivering power to the load.

The UPS operates in double-conversion (online) mode, where all the incoming power is continuously processed by the front-end rectifier and GFM stages. This configuration ensures high power quality and reliability for critical loads, but also makes the front-end rectifier a key component for maintaining stable operation. Any disturbance or instability at the rectifier stage directly propagates through the dc-link and affects the entire system.
In addition to the IT load, the model includes simplified cooling and auxiliary site-support loads to represent the overall facility structure. These components contribute to the total data center demand, while the IT load is modeled as the dominant dynamic component.

\subsection{UPS Internal Structure}
The centralized UPS system consists of a three-phase voltage-source rectifier, a common dc-link, a GFM, and a battery energy storage system interfaced through a bidirectional dc–dc converter. Under normal operation, power flows from the grid through the rectifier to the dc-link, which acts as the main energy interface between the grid-side and load-side converters.

The front-end rectifier plays a critical role in this architecture, as it governs the interaction between the data center and the upstream grid. In addition to regulating the dc-link voltage, the rectifier controls the exchange of active and reactive power with the grid, directly influencing system stability, particularly under weak-grid conditions.

In this work, the focus is placed on the control of the front-end rectifier. The GFM and dc–dc converter are modeled using conventional control schemes commonly adopted in converter-dominated power systems \cite{shah2021review,vasquez2022benchmarking} to ensure stable internal UPS operation and isolate the impact of the rectifier control strategy.

\subsection{Dynamic IT Load Modeling}
The load profile considered in this work represents the aggregated behavior of rack-level IT equipment, which constitutes the dominant portion of data center power consumption. This aggregated representation is consistent with grid-level modeling practices, where large numbers of individual devices are simplified into equivalent dynamic loads \cite{PNNL2025}.

\begin{figure}[h]
\centering
\includegraphics[trim={0.6cm 0.1cm 0cm 0.4cm},clip,width=0.97\columnwidth]{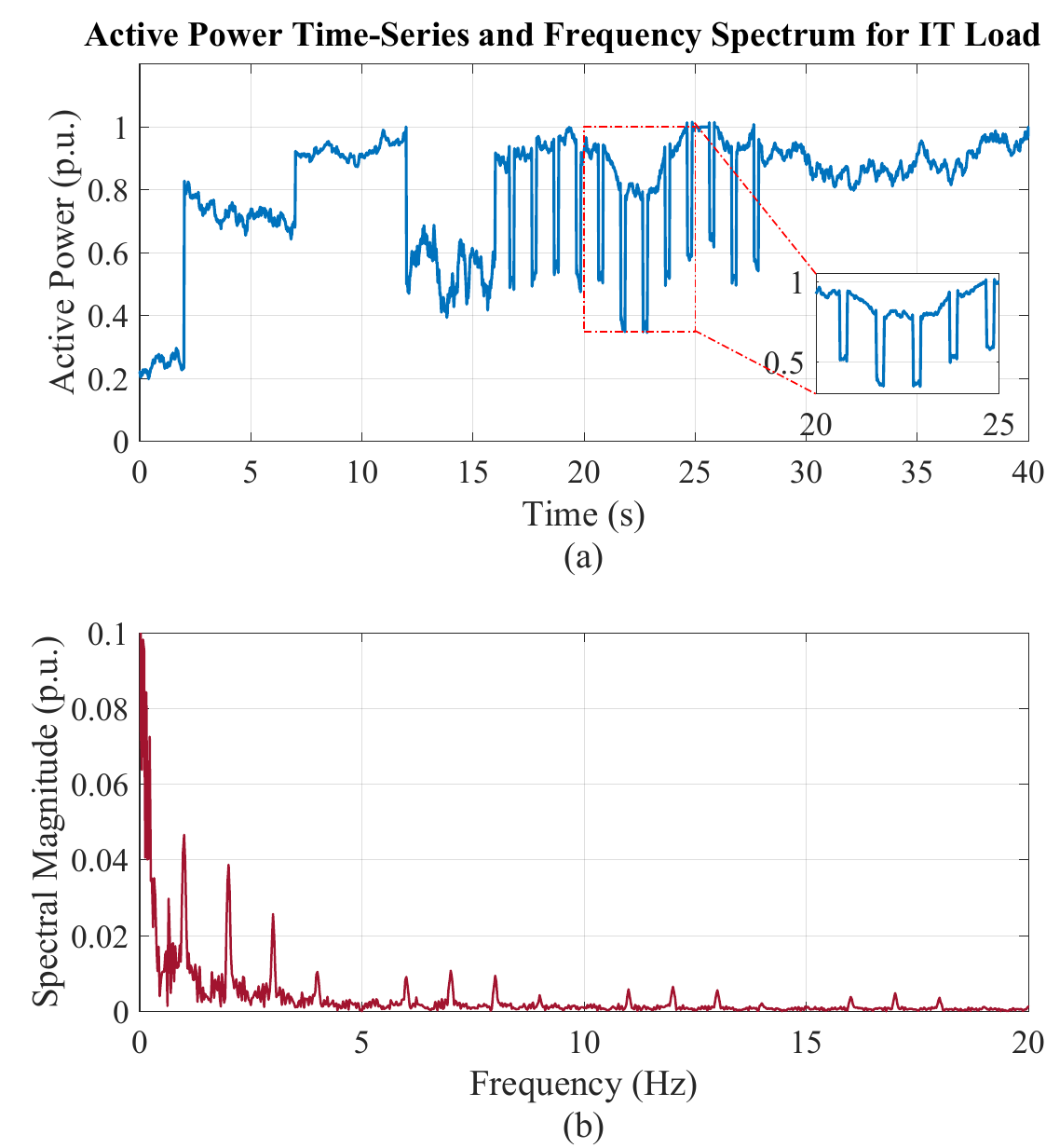}
\vspace{-1em}
\caption{Aggregated rack-level active power waveform and corresponding frequency spectrum. The spectrum reveals dominant low-frequency components.} 
\label{fig:IT_load_profile}
\vspace{-0.2cm}
\end{figure}

In modern AI-oriented data centers, synchronized compute and communication processes across large GPU clusters can introduce structured and correlated power variations \cite{choukse2025power}. Fig.~\ref{fig:IT_load_profile} shows the adopted active power waveform and its frequency spectrum. The waveform includes abrupt transitions between high and low power levels, while the spectrum reveals dominant low-frequency components primarily below 5 Hz. These components are relevant because they can interact with grid dynamics and converter control loops under weak-grid conditions.

The presence of low-frequency components in the load spectrum highlights the importance of capturing dynamic interactions between the load and the power electronic interface, particularly under weak-grid conditions. Therefore, the adopted load model provides a realistic representation of aggregated rack-level demand and serves as an effective excitation signal for evaluating the performance of the proposed control strategy.

\subsection{Weak-Grid Representation}

The upstream grid is modeled using a Thevenin equivalent representation to emulate different grid strength conditions at the point of connection of the data center. As illustrated in Fig.~\ref{fig:architecture}, the equivalent is defined at the 480 V node upstream of the UPS system, where the data center and other loads are connected to the distribution network.

The Thevenin equivalent consists of a voltage source and an equivalent impedance, which represents the aggregated effect of the upstream network. By adjusting this impedance, different grid strength conditions can be reproduced. In particular, the SCR is used to quantify grid strength, defined as the ratio between the grid short-circuit capacity and the rated power of the converter.
In this work, weak-grid conditions are emulated by selecting a low SCR value (e.g., SCR $\approx 2$), which is representative of converter-dominated systems with reduced system strength. Under such conditions, the interaction between grid impedance and converter control dynamics becomes significant, especially in the presence of dynamic loads.

Such modeling approach allows the evaluation of the interaction between the UPS system and the upstream grid under realistic operating conditions, where both the aggregated IT load and other facility loads are connected at the same node, as shown in Fig.~\ref{fig:architecture}.

\section{Rectifier Control Strategies}

\subsection{Conventional PI-Based Rectifier Control}
\begin{figure}[h]
\centering
\includegraphics[trim={0.6cm 0.2cm 0.6cm 0.7cm},clip,width=0.8\columnwidth]{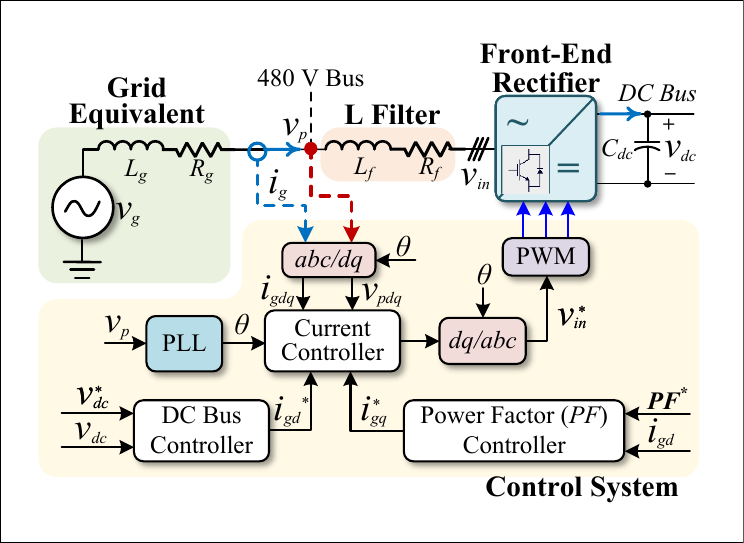}
\vspace{-1.8em}
\caption{Conventional PLL-based PI control structure of the front-end rectifier.} 
\label{fig:classical_control_rectifier}
\vspace{-1.2em}
\end{figure}

Fig.~\ref{fig:classical_control_rectifier} shows the conventional control structure used for the front-end rectifier of the centralized UPS system. The rectifier is connected to the 480 V bus through an L filter, while the upstream grid is represented by an equivalent voltage source and impedance to emulate weak-grid conditions.

The PLL extracts the phase angle from the point of common coupling (PCC)  voltage, which is used to transform measured three-phase quantities into dq reference frame. An outer dc-link voltage control loop generates the d-axis current reference, while the q-axis current reference is determined by a power factor controller, typically set to unity power factor operation~\cite{shah2021review}.
Inner current control loops regulate the d- and q-axis currents using PI controllers. Decoupling terms are included to compensate for cross-coupling effects, and the resulting voltage references are transformed back to the abc frame and applied to the PWM stage \cite{vasquez2022benchmarking}.

Such control strategy is widely used in industrial UPS systems due to its simplicity and effectiveness under stiff grid conditions. However, its performance depends on accurate synchronization and on the assumption of relatively stable grid voltages. Under the operating conditions defined in Section~II.C, these assumptions may not hold, which can affect system stability.

\subsection{VM-DPC-Based Rectifier Control}

The VM-DPC was proposed to regulate the active and reactive power to enhance the transient response and stability. The VM-DPC concept has been previously applied to a three-phase PWM rectifier \cite{Gui2018Rectifier} and later extended to weak grid applications~\cite{Gui2019VMDPC}. In this work, the method is applied to centralized UPS rectifiers supplying dynamic data center loads. 

The fundamental component of the PCC voltage is obtained as $v_{p\alpha\beta,f} = G_{bpf}(s)\,v_{\alpha\beta}$, where $G_{bpf}(s)$ is a band-pass filter centered at the fundamental frequency. This filtering stage is essential for reliable VM-DPC operation under weak-grid conditions, as it ensures accurate power calculation and prevents instability caused by voltage distortion \cite{Gui2019VMDPC}.

Using the filtered PCC voltage, the active and reactive powers are computed a:

\begin{equation}
P = \frac{3}{2}(v_{p\alpha,f} i_{g\alpha} + v_{p\beta,f} i_{g\beta}),
\end{equation}
\begin{equation}
Q = \frac{3}{2}(v_{p\beta,f} i_{g\alpha} - v_{p\alpha,f} i_{g\beta}),
\end{equation}
where $v_{p \alpha\beta,f}$ is the filtered PCC voltage component and $i_{g\alpha\beta}$ the grid-side current components.

For an L-filtered front-end rectifier, the power dynamics can be expressed as:
\begin{equation}
\dot{P} = -\frac{R_f}{L_f}P - \omega Q + \frac{3}{2L_f}(v_{p\alpha,f}u_{\alpha} + v_{p\beta,f}u_{\beta}) - \frac{3}{2L_f}V_p^2,
\end{equation}
\begin{equation}
\dot{Q} = \omega P - \frac{R_f}{L_f}Q + \frac{3}{2L_f}(v_{p\beta,f}u_{\alpha} - v_{p\alpha,f}u_{\beta}),
\end{equation}
where $u_\alpha$ and $u_\beta$ are the converter voltage components and $V_p^2 = v_{p\alpha,f}^2 + v_{p\beta,f}^2$.

To simplify the control design, VM-DPC defines the voltage-modulated inputs:
\begin{equation}
u_P = v_{p\alpha,f}u_{\alpha} + v_{p\beta,f}u_{\beta} - V_p^2,
\end{equation}  
\begin{equation}
u_Q = v_{p\beta,f}u_{\alpha} + v_{p\alpha,f}u_{\beta}.
\end{equation}
With voltage-modulated inputs, the power dynamics become:
\begin{equation}
\dot{P} = -\frac{R_f}{L_f}P - \omega~Q + \frac{3}{2L_f}u_P,
\end{equation}
\begin{equation}
\dot{Q} = -\omega~P -\frac{R_f}{L_f}Q + u_Q.
\end{equation}

To cancel the power coupling terms and yields stable error dynamics for positive controller gains, the control inputs are then selected as:
\begin{equation}
u_P = \frac{2L\omega~Q}{3} +  K_{p,P}(P^{\ast} - P) + K_{i,P}\int (P^{\ast} - P)dt,
\end{equation}
\begin{equation}
u_Q =  -\frac{2L\omega~P}{3} + K_{p,Q}(Q^{\ast} - Q) + K_{i,Q}\int (Q^{\ast} - Q)dt,
\end{equation}
where $K_{p,P}-K_{i,P}$ and 
$K_{p,Q} - K_{i,Q}$ are the proportional and integral gains of the controllers regulating the active power $P$ and reactive power $Q$, respectively.

The converter voltage references are recovered from the inverse voltage modulation as:
\begin{equation}
u_{\alpha}^{\ast} = \frac{v_{p\alpha,f}(u_P + V_p^2) - v_{p\beta,f}u_Q}{V_p^2},
\end{equation}
\begin{equation}
u_{\beta}^{\ast} = \frac{v_{p\beta,f}(u_P + V_p^2) + v_{p\alpha,f}u_Q}{V_p^2}. 
\end{equation}

These voltage references are transformed to three-phase commands and applied to the PWM stage.

In centralized data center UPS systems, the active power reference $P^*$ is controlled by the IT load demand through dc-link regulation, reflecting the highly dynamic and bursty nature of AI workloads. Since the same power sign convention as the VM-DPC formulation is used, rectifier operation corresponds to drawing active power from the grid toward the dc-link.

In this work, the control strategy is extended by explicitly leveraging reactive power control to enhance stability under rapid IT load variations. Specifically, the rectifier is operated such that active power is drawn from the grid to supply the IT load, while a constant reactive power reference $Q^*$ is injected into the grid to support the PCC voltage.

This operating condition is particularly relevant in data centers, where large and synchronized load changes can induce low-frequency power oscillations. By injecting reactive power, the rectifier contributes to stabilizing the PCC voltage and improving damping under weak-grid conditions.
This modification represents a key contribution of this work, as it adapts the VM-DPC framework to the unique characteristics of data center power systems.

\section{Simulation Results}

\subsection{Simulation Setup and Test Cases}
The performance of the centralized UPS data center system is evaluated using detailed EMT simulations under different grid strength conditions. The objective is to assess the stability of the overall data center power system when subjected to highly dynamic IT load variations and weak-grid conditions.

The simulated system corresponds to the centralized UPS data center architecture described in Section II, where the grid-side rectifier, dc-link, GFM, and downstream loads operate as a tightly coupled system. In this configuration, the rectifier is not an isolated component but a critical interface that governs the interaction between the upstream grid and the internal data center power infrastructure. The upstream grid is modeled using a Thevenin equivalent, and the SCR is varied to emulate different grid strengths. Two cases are considered: moderate grid strength ($SCR \approx 3$)  and weak-grid condition ($SCR\approx2$). The same data center model and IT load profile are used in all simulations to ensure a consistent comparison between control strategies.

The data center loading is defined based on a power usage effectiveness (PUE) of 1.3. The IT load corresponds to approximately 0.769 MVA, while the remaining power demand is attributed to cooling systems and auxiliary (site support) loads. This results in a total facility demand of approximately 1 MVA. While the IT load is modeled as a highly dynamic power profile, the cooling and auxiliary loads are represented as slower-varying components that primarily contribute to the steady-state power demand.

This distinction is important because the IT load dominates the fast power variations, which are directly processed by the UPS system. As a result, any instability in the rectifier stage propagates through the dc-link and affects the entire data center power system, including the GFM and the supplied loads.

\begin{table}[h]
\centering
\caption{Main Simulation Parameters for the Data Center Model}
\label{tab:simulation_parameters}
\renewcommand{\arraystretch}{1.15}
\begin{tabular}{lll}
\hline
\textbf{Category} & \textbf{Parameter} & \textbf{Value} \\
\hline

\multicolumn{3}{c}{\textit{Grid and Facility Parameters}} \\
\hline
Grid voltage & $V_{g_{LL,rms}} $ & 480 V \\
Grid frequency & $f_g$ & 60 Hz \\
Short-circuit ratio & SCR & 3, 2 \\
Grid resistance & $R_g$ & 0.02, 0.04 $\Omega$ \\
Grid inductance & $L_g$ & 0.15, 0.21 mH\\
Data center PUE & PUE & 1.3 \\
Total facility rating & $S_{\mathrm{facility}}$ & 1 MVA \\
IT load rating & $S_{\mathrm{IT}}$ & 0.77 MVA \\
Cooling + support load & $S_{\mathrm{aux}}$ & 0.23 MVA \\
\hline

\multicolumn{3}{c}{\textit{Front-End Rectifier Parameters}} \\
\hline
Rated Power & $S_{f}$ & 1.2 MVA \\
dc-link voltage & $v_{dc}$ & 1600 V \\
DC-link capacitance & $C_{dc}$ & 22.5 mF \\
Filter inductance & $L_f$ & 0.5 mH \\
Filter resistance & $R_f$ & 0.01 $\Omega$ \\
Switching frequency & $f_{sw}$ & 10 kHz \\
\hline

\multicolumn{3}{c}{\textit{Grid-Forming Inverter Parameters}} \\
\hline
Rated power & $S_{\mathrm{gfm}}$ & 1 MVA \\
Output voltage & $V_{load_{rms}}$ & 240 V \\
Filter inductance & $L_{gfm    }$ & 0.15 mH \\
Filter resistance & $R_{gfm}$ & 0.001 $\Omega$ \\
Filter capacitance & $C_{gfm}$ & 1.5 mF \\
\hline


\multicolumn{3}{c}{\textit{Conventional PI Rectifier Control}} \\
\hline
dc-link proportional gain & $K_{p,dc}$ & 8.6 \\
dc-link integral gain & $K_{i,dc}$ & 40.6 \\
Current loop proportional gain & $K_{p,i}$ & 15.2 \\
Current loop integral gain & $K_{i,i}$ & 86 \\
Power factor reference & $PF^{\ast}$ & 1.0 \\
\hline

\multicolumn{3}{c}{\textit{VM-DPC Rectifier Control}} \\
\hline
Proportional gain & $K_{p,P}/K_{p,Q}$ & 2500 \\
Integral gain & $K_{i,P}/K_{i,Q}$ & 25000 \\
BPF center frequency & $f_{bpf}$ & 60 Hz \\
BPF damping ratio & $\zeta_{bpf}$ & 0.707 \\
\hline


\end{tabular}
\vspace{-2em}
\end{table}

Two control strategies are evaluated at the front-end rectifier. Conventional PI-based control with PLL synchronization and the VM-DPC  without PLL. For the VM-DPC implementation, the active power reference is dictated by the dc-link voltage regulation, while a constant reactive power reference is injected into the grid. Different values of $Q^*$ are evaluated to analyze their impact on system stability.

The main system parameters, including electrical ratings, controller settings, and load characteristics, are summarized in Table I.

\subsection{Baseline Operation Under Moderate Grid Strength}
\begin{figure}[h]
\centering
\includegraphics[trim={1.05cm 0cm 0cm 0cm},clip,width=0.95\columnwidth]{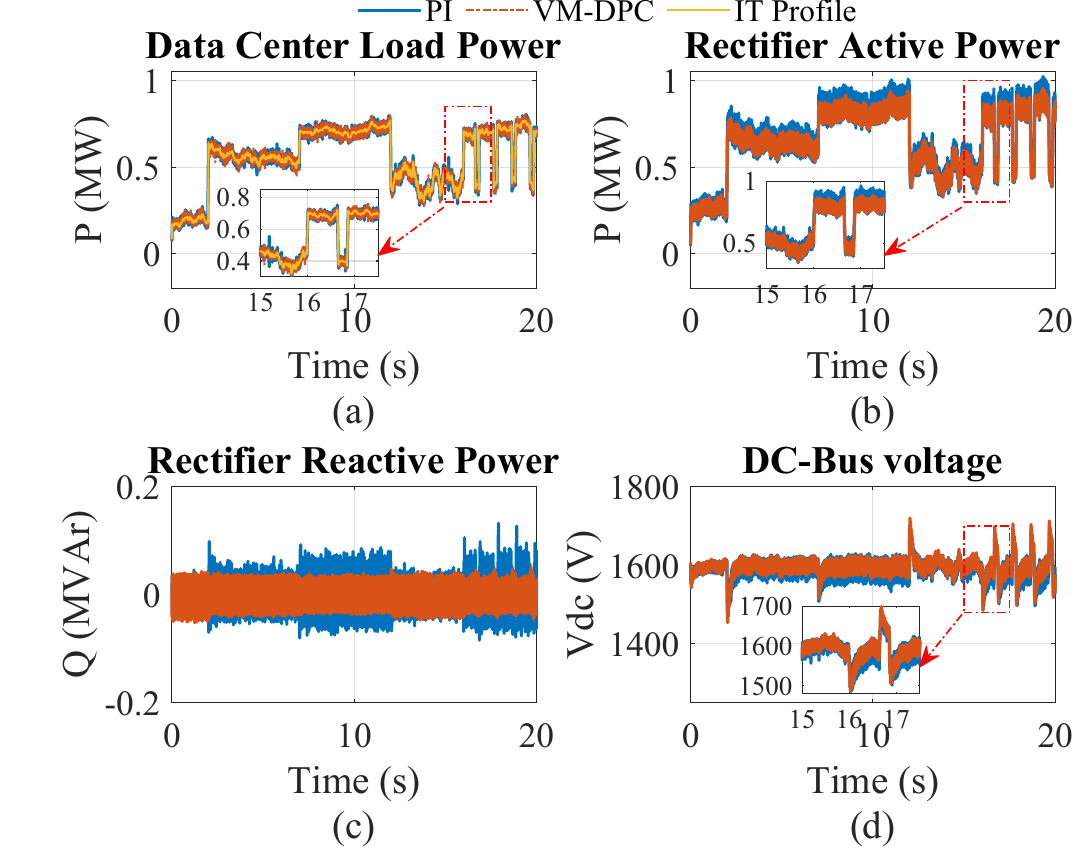}
\vspace{-1em}
\caption{Baseline response of the centralized UPS data center system under $SCR \approx 3$ comparing PI control and VM-DPC}
\label{fig:baseline}
\end{figure}

The baseline performance of the centralized UPS data center system is evaluated under moderate grid strength conditions ($SCR \approx 3$). Fig.~\ref{fig:baseline} compares the response of the conventional PI-based controller and the VM-DPC approach under the same dynamic IT load profile

Fig.~\ref{fig:baseline}(a) shows the active power delivered by the UPS to the downstream data center loads together with the IT load profile. Both control strategies successfully supply the dynamic load demand while maintaining stable operation. The corresponding rectifier active and reactive powers are shown in Fig.~\ref{fig:baseline}(b) and Fig.~\ref{fig:baseline}(c), respectively. In the VM-DPC case, reactive power is injected toward the grid to provide voltage support and improve the interaction between the UPS system and the upstream network.

The dc-link voltage response is presented in Fig.~\ref{fig:baseline}(d). Both controllers maintain the dc-link voltage within acceptable limits, ensuring stable operation of the GFM and continuous power delivery to the IT infrastructure. However, the VM-DPC controller exhibits slightly improved damping during abrupt load variations.

Overall, the results confirm that both control strategies can maintain stable operation under moderate grid strength conditions. More importantly, this baseline case shows that the dynamic IT load profile alone does not destabilize the system; instability emerges when these fast load variations interact with weak-grid conditions.

\subsection{Comparison Under weak-grid Conditions}
\begin{figure}[h]
\centering
\includegraphics[trim={0.8cm 0cm 0cm 0cm},clip,width=\columnwidth]{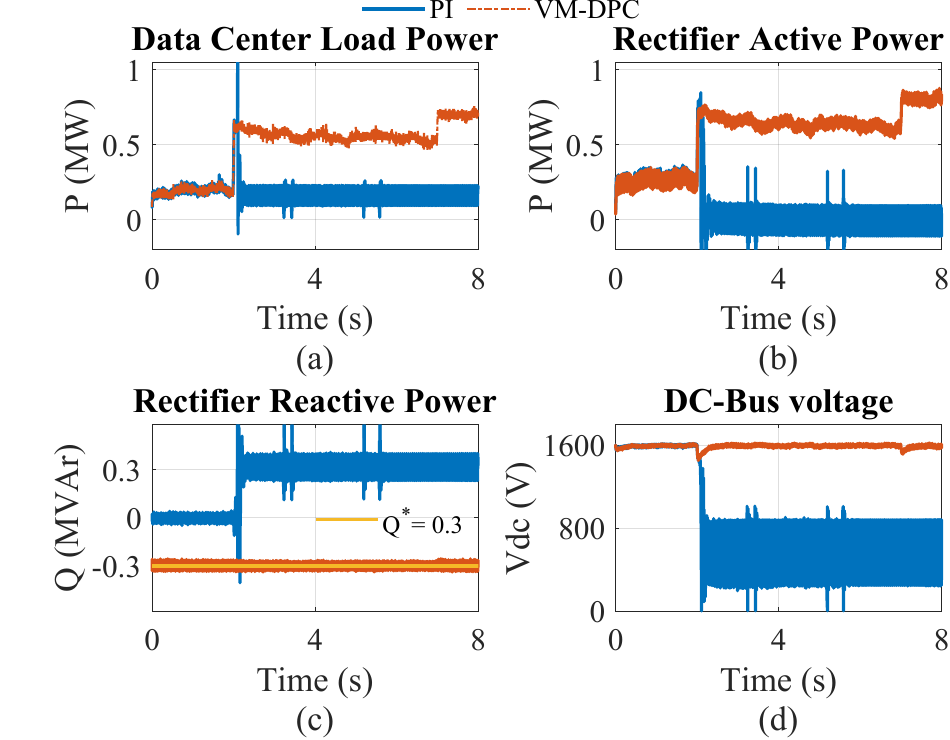}
\vspace{-2em}
    \caption{Comparison between conventional PI control and VM-DPC under weak-grid conditions ($SCR \approx 2$).}
\label{fig:weakgrid}
\end{figure}

The performance of the centralized UPS data center system under weak-grid conditions is evaluated using an SCR of approximately 2. Fig.~\ref{fig:weakgrid} compares the response of the conventional PI-based controller and the VM-DPC approach following a large increase in the IT load demand at $t\approx2s$.

As shown in Fig.~\ref{fig:weakgrid}(a) and Fig.~\ref{fig:weakgrid}(b), the PI-controlled system becomes unstable after the load transition and loses the ability to sustain stable power delivery to the data center loads. The instability originates from the interaction between the PLL dynamics and the weak-grid impedance, leading to severe oscillations in the rectifier active power.

The corresponding reactive power response is shown in Fig.~\ref{fig:weakgrid}(c). In the VM-DPC case, reactive power is injected toward the grid to provide voltage support and improve the interaction between the UPS system and the upstream network. As a result, the VM-DPC controller maintains stable operation and a well-damped response despite the reduced grid strength.

The dc-link voltage response is presented in Fig. 5(d). While the PI-controlled system experiences dc-link voltage collapse following the load variation, the VM-DPC approach maintains the dc-link voltage within acceptable operating limits, ensuring continuous and stable power delivery to the IT infrastructure.

Overall, the results demonstrate that VM-DPC significantly improves the stability of centralized UPS data center systems under weak-grid conditions and highly dynamic IT load variations.

\subsection{Stability Dependence on Reactive Power Injection}
\begin{figure}[h]
\centering
\includegraphics[trim={1.2cm 0cm 0cm 0cm},clip,width=\columnwidth]{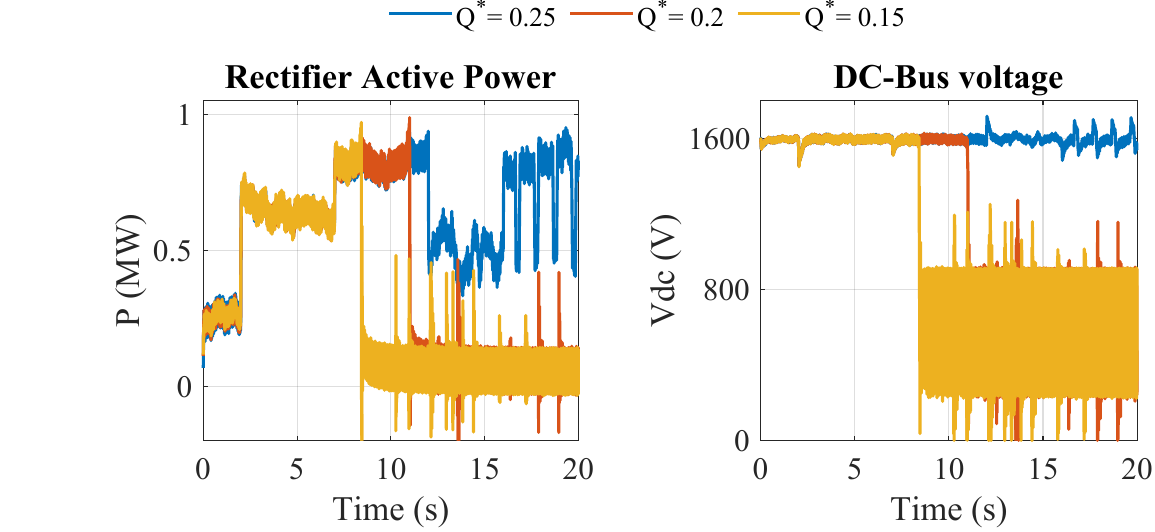}
\vspace{-2em}
\caption{Effect of reactive power support on VM-DPC stability under weak-grid conditions ($SCR \approx 2$).}
\label{fig:VMDPC_VariationsQ}
\end{figure}

Fig.~\ref{fig:VMDPC_VariationsQ} evaluates the effect of reactive power support on the stability of the VM-DPC-controlled centralized UPS system under weak-grid conditions ($SCR\approx2$). Different fixed reactive power references are considered while maintaining the same dynamic IT load profile.
\vspace{-1em}

As shown in Fig.~\ref{fig:VMDPC_VariationsQ}, insufficient reactive power support leads to instability following the large load transition. For lower values of $Q^*$, the front-end rectifier active power becomes unstable and the dc-link voltage collapses after the disturbance. In contrast, increasing the reactive power injection improves system damping and maintains stable operation.
 
Among the evaluated cases, $Q^*=0.25$~MVAr provides stable operation throughout the simulation, maintaining both stable power transfer and acceptable dc-link voltage regulation. These results demonstrate that reactive power support plays a critical role in maintaining the stability of centralized UPS data center systems operating under weak-grid conditions and highly dynamic IT load variations.

\section{Conclusion}

This paper presented the application of VM-DPC to the front-end rectifier of a centralized UPS data center system operating under the weak-grid condition and highly dynamic AI-oriented IT load variations. A detailed EMT model was used to evaluate the interaction between the UPS system, the dynamic IT load, and the upstream weak-grid.
The results showed that conventional PLL-based PI control becomes unstable under the weak-grid condition ($SCR\approx2$) following large IT load variations, resulting in dc-link voltage collapse and loss of stable power delivery. In contrast, the VM-DPC approach maintained stable operation through reactive power support to the grid.
Additionally, the results demonstrated that system stability strongly depends on the injected reactive power level, where adequate reactive power support significantly improves system damping and dc-link voltage regulation. Overall, the presented results highlight the potential of VM-DPC as a grid-supportive control strategy for future centralized UPS data center systems operating under weak-grid conditions.

\bibliographystyle{IEEEtran}
\bibliography{References} %

\end{document}